\documentclass[runningheads]{llncs}
\usepackage[T1]{fontenc}
\usepackage[utf8]{inputenc}
\usepackage{verbatim}
\usepackage{amsmath}
\usepackage{dsfont}
\usepackage{pgf-umlsd}
\usepackage{hyperref}
\usepackage{graphicx}
\begin{document}
\title{Stream Containers for Resource-oriented RDF Stream Processing\thanks{This work was funded by the German Federal Ministry of Education and Research through the MOSAIK project (grant no. 01IS18070A).}}
%
%
\author{Daniel Schraudner\orcidID{0000-0002-2660-676X}\textsuperscript{1} \and
Andreas Harth\orcidID{0000-0002-0702-510X}\textsuperscript{1,2}}
\authorrunning{D. Schraudner \and A. Harth}
%
\institute{\textsuperscript{1}Chair of Technical Information Systems, Friedrich-Alexander-University Erlangen-Nürnberg, Nuremberg, Germany\\
\email{\{daniel.schraudner,andreas.harth\}@fau.de}\\
\textsuperscript{2}Fraunhofer IIS, Fraunhofer Institute for Integrated Circuits IIS, Nuremberg, Germany}
\maketitle              
\begin{abstract}
We introduce Stream Containers inspired by the Linked Data Platform as an alternative way to process RDF streams. A Stream Container represents a single RDF data stream that can be accessed in a resource-oriented way which allows for better interoperability with the existing Semantic Web infrastructure. Stream Containers are managed by webservers that are responsible for implementing the S2R operator, i.e. calculating the window for their clients. The clients on the other hand can use a standard SPARQL processor in combination with HTTP requests to do RDF processing. Query results can be converted back to an RDF stream (R2S operator) by posting the data to a Stream Container. Our approach of resource-oriented RDF stream processing can lead to a better distribution of load and thus to better worldwide scalability. We give a general overview of the proposed architecture as well as the formal semantics of the overall system.

\end{abstract}
\section{Introduction}
\enlargethispage{1cm}
With more and more data being available today in very short intervals processing streams of data -- in contrast to the traditional mode of batch processing -- is becoming more important \cite{10.5555/2823980}. Stream processing allows data to be handled in (near) real-time which is advantageous for modern applications like Big Data and Industry 4.0 that need to process big amounts of data fast.

The Semantic Web, originally being a static collection of data, is influenced by this development, too. The field of RDF stream processing (RSP) has brought forth several extensions to the SPARQL query language (e.g. \cite{barbieri2010c}, \cite{10.1007/978-3-642-17746-0_7}, \cite{10.1007/978-3-642-25073-6_24}) that allow to also query streams of RDF graphs and triples.

However, current RSP implementations, even though being considered Semantic Web technologies, are not really aligned to the architecture of the Web. Existing systems instead are monolithic pieces of software that receive one or several RDF streams as input and on the other side push out streams of (RDF) data. In the traditional Web architecture which can be characterized by the Representational State Transfer (REST) architectural style \cite{fielding2000architectural}, there exists a clear separation of concerns between clients and servers. Clients can retrieve data from the server using e.g. an HTTP GET request, process the data and then send it back to a webserver using e.g. an HTTP PUT request. For current RSP implementations it is not clear whether the query engines are clients or servers and how they exchange data with others. In fact, they can instead better be compared to a publish-subscribe architecture that is used by Internet of Things protocols like MQTT (which follows a push instead of a pull model).

The misalignment of existing RSP implementations with the Web architecture poses two problems: First, the (Semantic) Web is built to scale in a worldwide fashion but using a push instead of a pull model is infeasible for web-scale applications \cite{fielding2000architectural}. It will not be possible to use current implementations in an application that is distributed over the whole world -- as opposed to an implementation conforming to REST and thus using the pull model where caching, clustering and load-balancing allow the application to scale to a very large extend \cite{pautasso2008restful}.

The second problem is related to the first one: An RSP application that wants to scale worldwide needs to be easily interfaceable with other systems. For RSP it is still an unsolved issue, how to exchange RDF streams among different types of RSP systems \cite{dell2017web}. If properly aligned to the web architecture, those systems could, even if they are very heterogeneous, easily exchange streams with each other and also with the existing (Semantic) Web infrastructure (like  e.g. Linked Data, one of the main applications of the Semantic Web, which is also build upon the REST architectural style).

The main challenge when aligning RSP with the Web architecture is that at first sight, it seems very unnatural to do this at all. A stream, naively thought, is data that continuously is created somewhere and then immediately needs to be sent to where it is to be consumed; using a push model would be the obvious thing to do. Nevertheless, because of the advantages of an alignment with the Web architecture mentioned before, we think that it is very promising to map RDF streams to a RESTful architecture. 

For this purpose, we propose \textit{Stream Containers} as an extension to the Linked Data Platform \cite{speicher2015linked}. Stream Containers are a resource-oriented implementation of RDF Streams. They are managed by a webserver and responsible for transforming the stream into a set of RDF graphs (windowing operators) when requested by a client. The client can then implement the actual processing of the data e.g. using a standard SPARQL query processor. When new data (in the form of RDF graphs) is pushed to the Stream Container, the webserver is also responsible for transforming it into an RDF stream (streaming operators). This decoupling between client and server will allow us to spawn as many Stream Containers and processing clients as we want. Because we are building upon an established RESTful Linked Data protocol, the Stream Containers and clients will all be interoperable with each other and the existing Web infrastructure. All without needing to extend or change the established RDF data model.

All in all, we propose a novel approach to do RDF stream processing in a Web-scalable and interoperable way and give an overview of the intended system architecture. We furthermore provide a formalization of the computations done by the Stream Container as well as the semantics of our overall system.
\enlargethispage{1cm}

The remainder of this work is structured as follows: In Section~\ref{sec:related_work} we give an overview of existing approaches to RDF stream processing and how they relate to our approach. In Section~\ref{sec:approach} we introduce our Stream Container approach. Therefore we first outline the general indented architecture, explain how Stream Containers are implemented and give the formal semantics of the system. Finally, in Section~\ref{sec:conclusion} we give a conclusion and an outlook on our future work.

\section{Related Work}
\label{sec:related_work}

There already exists a lot of proposals for systems that do RDF stream processing, most of them being extensions to the SPARQL query language. We will go into detail about the three most noted approaches.

\paragraph{C-SPARQL} \cite{barbieri2010c} is an extension to SPARQL where RDF streams are represented as a sequence of RDF triples together with timestamps. C-SPARQL allows specifying physical windows (the $n$ most recent triples) and logical windows (all triples with a timestamp more recent than $t$). For queries to operate on an RDF stream they have to be registered as continuous queries with the C-SPARQL engine. The results of those queries are transformed back into a stream. The timestamp of the triples can be used in the query by utilizing a built-in C-SPARQL function. A formal specification of the syntax, the semantics and some examples are provided, however it is not exactly clear, how to interface a C-SPARQL query processor with other RDF stream processors or even with arbitrary Linked Data. Internally the C-SPARQL engine uses a data stream management system to decompose continuous C-SPARQL queries into several SPARQL queries and a standard SPARQL engine to process those queries.
\enlargethispage{1cm}







\paragraph{SPARQL$_{Stream}$} \cite{10.1007/978-3-642-17746-0_7} is similar to C-SPARQL in that it provides an extension to SPARQL to handle RDF streams, where streams are also defined as a sequence of triple-timestamp tuples. In SPARQL$_{Stream}$ continuous queries also have to be registered with the query engine, but the queries are transformed (using stream-to-ontology-mappings) into queries in the relational streaming query language SNEEql (which means it is kind of the dual approach to C-SPARQL). SPARQL$_{Stream}$ only supports logical windows but has support for different streaming operators like \texttt{RSTREAM}, \texttt{ISTREAM} and \texttt{DSTREAM}.







\paragraph{CQELS} \cite{10.1007/978-3-642-25073-6_24} differs from the previous two approaches in that it does not do any translation of queries to existing languages but instead provides a native implementation for the query language. The authors call this a white-box approach which allows better optimization of the system and thus leads to a better performance than the other systems. CQELS has physical and logical window operators available but only provides the standard streaming operator \texttt{RSTREAM}.






\paragraph{RSP-QL} \cite{dell2014rsp} is a formal model that has been proposed in 2014 to unify existing approaches to RDF stream processing (especially those introduced here) and tries to explain their heterogeneity. The authors provide a formal semantics that subsumes existing implementations and compare the features of different implementations against each other.

However, all of the approaches introduced until now, suffer from the problems described in our Introduction: These RSP query engines are monolithic pieces of software with no existing decoupling between a client that does the processing work and a server that provides the data, which is required to be properly aligned with the Web architecture. Our approach on the other side tries to tackle the said issues while still maintaining the same execution semantics like e.g. stated by RSP-QL.









\paragraph{TPF-QS} \cite{taelman2018semantics} extends the Triple Pattern Fragment model to be used for stream processing. A clear separation between client and server like in our work exists, in this case to minimize the load on the server-side. The client accesses the stream by pulling, however the windowing operator is calculated at the client-side, leading to an increased usage of bandwidth which can be a limiting factor.

\paragraph{TripleWave} \cite{DBLP:conf/semweb/0001CDB0VA16} is a framework for publishing RDF streams on the Web. It allows to either use existing non-RDF streams or time-annotated RDF, that is not in the form of a stream, as input. On the other side, TripleWave allows to output streams either in a push or a pull-based fashion. This way existing RSP engines can be easily used together with TripleWave which is the first step towards interoperability between RSP engines, however this is just an adapter between different implementations, the problem of missing worldwide scalability by decoupling stream and stream processing is not considered.

\paragraph{Linked Data Notifications.} In his work Calbimonte \cite{calbimonte2017linked} proposes the idea to use Linked Data Notifications (LDN) \cite{capadisli2017linked} to represent RDF streams. Like our Stream Containers, Linked Data Notifications are also based on the Linked Data Platform standard. Also, the general idea of our approach, using Web resources to represent RDF streams and their data, is very similar to the LDN approach. Even though this is a step towards the right direction -- namely using the Web architecture for better scalability as well as better integration and inter-operation -- the author stays quite vague about how his idea can be concretely realized.

\section{Approach}
\label{sec:approach}
\subsection{Architecture}
In RDF stream processing there usually exist three types of operators to handle RDF streaming and non-streaming data. In reference to the (non-RDF) stream processing language CQL \cite{arasu2006cql}, we call the first type of operators S2R (or windowing operators). They are responsible for transforming an RDF stream into a set of static RDF graphs. We can do arbitrary processing on these static graphs using the R2R operators. To transform static RDF back into one or multiple RDF streams we can use the R2S (or streaming) operators.
\enlargethispage{1cm}

In the implementations, we discussed in our related work section, the S2R operators are implemented by specifying a window for the continuous query (e.g. using \texttt{WINDOW} and \texttt{STEP}), the R2R operators by using the standard SPARQL operators and the R2S operators by specifying a streaming behavior (e.g. \texttt{RSTREAM}) when registering the query.

These three types of operators are all implemented in one single monolithic RSP query processor. In contrast to this, we propose an architecture for RSP where the conversion between streaming and static RDF (S2R and R2S) is implemented on the server and thus decoupled from the actual processing of the static RDF, which happens on the clients -- plural as we can in a web-like manner almost arbitrarily scale up the number of clients doing processing work on the same stream.

The centerpiece of our proposal are Stream Containers that are managed by a webserver. They are responsible for maintaining the stream data, i.e. we can say that they represent an RDF stream. Furthermore, Stream Containers are responsible for computing the R2S and S2R operators by inserting a new data element (RDF graph) into the stream when it arrives through a POST request or by calculating the current window and returning the data elements in the window for every GET request respectively.

\begin{figure}
    \centering
    \includegraphics[width=\linewidth]{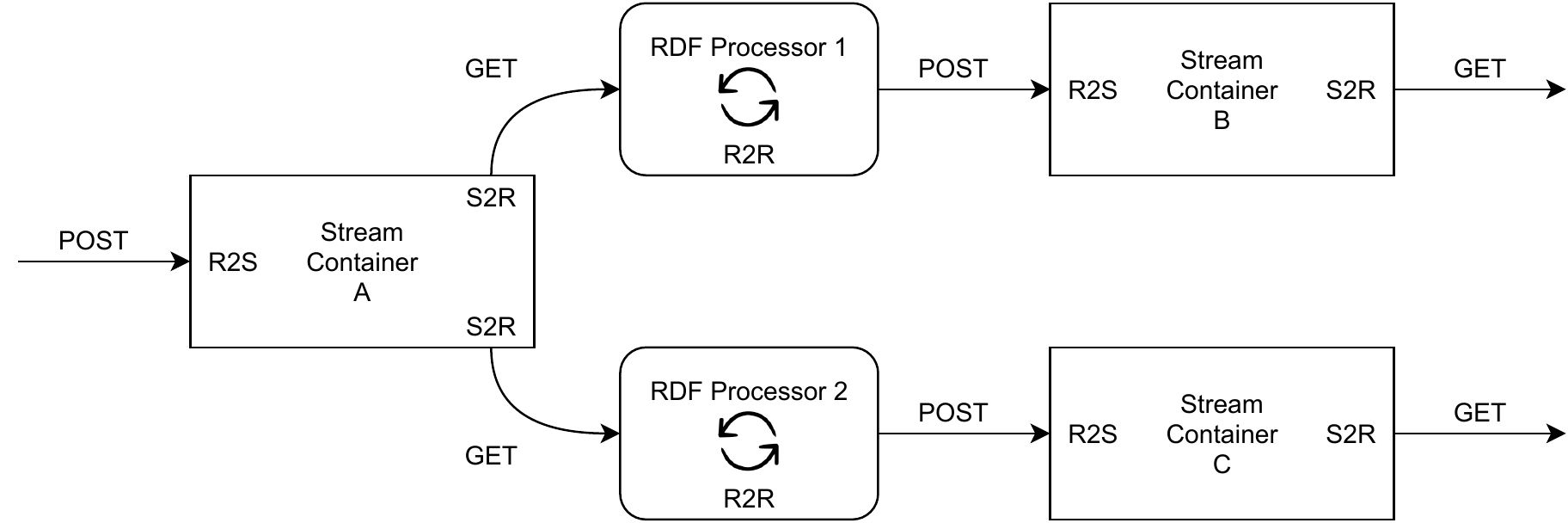}
    \caption{Architecture of a decoupled RSP system using Stream Containers}
    \label{fig:arch_sc}
\end{figure}

The general structure of the architecture we propose can be seen in Figure~\ref{fig:arch_sc}\footnote{Note that the arrow direction represents the direction of the data flow; the GET requests are of course not initiated by the Stream Containers}. The example comprises three distinct Stream Containers each managing a stream of RDF graphs. New graphs are continuously appended to \textit{Stream Container A} by POST requests from an external source on the left (R2S operators).

\textit{RDF Processors 1 \& 2} are continuously pulling the current content of the windows they specified at \textit{Stream Container A} (S2R operators), and processing the contents according to the R2R operators they are executing. Note that we are totally free to use whatever client implementations we want as long as they have an RDF an HTTP interface. We could e.g. use a simple standard SPARQL processor that is capable of doing HTTP requests to transform RDF graphs using SPARQL CONSTRUCT queries or a more sophisticated Linked Data processing system like Linked Data-Fu\footnote{https://linked-data-fu.github.io/} that is also capable of doing rule-based forward-chaining reasoning.

The result of the processing of \textit{RDF Processors 1 \& 2} can be transformed back into a stream by POSTing the result to an arbitrary stream container from where the stream data can again be retrieved by other entities.

\begin{figure}
    \centering
    \includegraphics[scale=0.68]{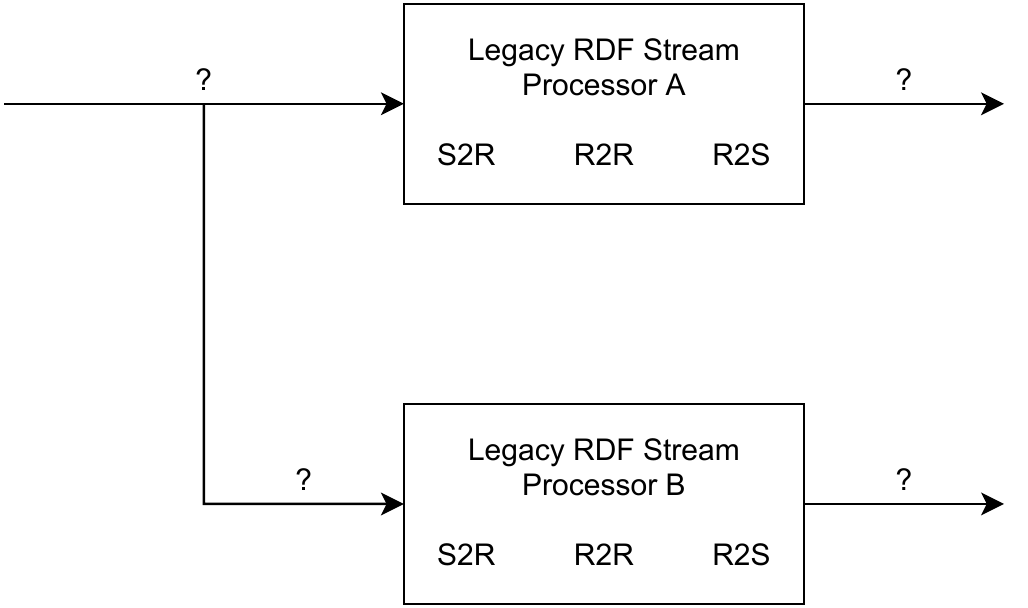}
    \caption{Architecture of a RSP system using existing stream processor implementations}
    \label{fig:arch_legacy}
\end{figure}

In contrast, the architecture of existing RDF Stream Processors can be seen in Figure~\ref{fig:arch_legacy}. We assume that both processors again are operating on the same RDF stream, however it is not exactly clear, how the data gets pushed to and from the stream processors (especially as the same data is used twice here). This problem has already been identified in \cite{dell2017web} and hinders interoperability between different RDF stream processing systems as well as existing Linked Data systems that are already in use.

\subsection{Stream Containers}
The Linked Data Platform (LDP) \cite{speicher2015linked} is a W3C recommendation to ease RESTful communication for Linked Data. LDP introduces the concept of containers, Web resources that can contain other Web resources (which can again be containers). By nesting containers into each other one can create a file system-like tree structure. According to the LDP protocol, Containers and their contained resources can be read using GET requests and manipulated using PUT and DELETE requests. New resources can be created by POSTing to an existing Container which creates a new resource with the payload of the request being contained in the addressed container.

LDP, as it is now, differentiates between three types of containers, \textit{Basic Containers}, \textit{Direct Containers} and \textit{Indirect Containers} where Basic Containers only provide the \textit{containment triples} (the triples stating which resources are contained by a Container) managed by the server. Direct and Indirect Containers add additional functionality by providing so-called \textit{membership triples}, generated by applying fixed rules to the contained resources\footnote{Direct Containers' membership triples relate a given resource to the contained resources with a given predicate, Indirect Containers' membership triples add a further indirection by relating a given resource to the objects of specified triples by a given predicate}.

We propose to extend the LDP model by \textit{Stream Containers}. Stream Containers maintain the data of an RDF stream by representing every data point of the stream as an RDF resource contained in the Stream Container. The contained resources are RDF graphs that have to include a triple that represents a timestamp\footnote{It is a common practice in the RSP world to separate the timestamp from the actual RDF graph, however we see no reason to do this, as the RDF data model is perfectly capable of expressing timestamps (for a further discussion see \cite{rula2012diversity})}.

So far the functionality is similar to a Basic Container but Stream Containers (resources with type \texttt{ldpsc:StreamContainer}\footnote{The Stream Container specific vocabulary is available at \href{https://solid.ti.rw.fau.de/public/ns/stream-containers}{https://solid.ti.rw.fau.de/public/ns/stream-containers}}) let us specify windows (e.g. by changing the RDF of the Container using a PUT request) using the \texttt{ldpsc:window} predicate. Windows are resources that have to specify a \texttt{ldp:has\-MemberRelation} and a \texttt{ldp:membershipResource}. The two properties work exactly the way they work for Direct and Indirect Containers (i.e. they specify the subject and predicate of the membership triples to be added). Additionally a window 
needs a \texttt{ldpsc:contentTimestampRelation} which specifies the predicate to look for timestamps inside of the contained resources (similar to \texttt{ldp:insertedContent\-Relation} for Indirect Containers) and either
\begin{itemize}
    \item 
    a time duration (\texttt{xsd:duration}) using \texttt{ldpsc:\-logical} which specifies the size of a logical window, or
    \item
    a number of triples (\texttt{xsd:integer}) using \texttt{ldpsc:physical} which specifies the size of a physical window.
\end{itemize} 
A Stream Container can have arbitrarily many different windows associated with it. Each of the windows can be manipulated by the clients independently and each window is responsible for materializing its membership triples individually.

\begin{figure}
\begin{verbatim}
@prefix ldpsc: <https://solid.ti.rw.fau.de/public/ns/stream-containers#> .
@prefix ldp: <http://www.w3.org/ns/ldp#> .
@prefix sosa: <http://www.w3.org/ns/sosa/> .
@prefix xsd: <http://www.w3.org/2001/XMLSchema#> .
@prefix ex: <http://example.org/> .

<> a ldpsc:StreamContainer ;
    ldp:contains </0> ,
                 </1> ,
                 </2> ,
                 </3> ,
    ldpsc:window [
        ldp:hasMemberRelation ex:inWindow ;
        ldp:membershipResource <#window1> ;
        ldpsc:contentTimestampRelation sosa:resultTime ;
        ldpsc:logical "PT2M"^^xsd:duration
    ] .

<#window1> ex:inWindow </2> ,
                       </3> .
\end{verbatim}
    \caption{Turtle representation of an exemple Stream Container}
    \label{fig:turtle_sc}
\end{figure}

\begin{figure}
\begin{verbatim}
@prefix sosa: <http://www.w3.org/ns/sosa/> .
@prefix xsd: <http://www.w3.org/2001/XMLSchema#> .
@prefix ex: <http://example.org/> .

</3> a sosa:Observation ;
    sosa:observedProperty ex:temperature ; 
    sosa:hasSimpleResult 22.3 ;
    sosa:resultTime "2021-07-20T10:51:08.657Z"^^xsd:dateTimeStamp .
\end{verbatim}
    \caption{Turtle representation of an example resource contained in a Stream Container}
    \label{fig:turtle_res}
\end{figure}

An example of a Stream Container can be seen in Figure~\ref{fig:turtle_sc}. The container contains four resources (\texttt{:0}, \texttt{:1}, \texttt{:2}, \texttt{:3}), representing RDF graphs in the stream which look e.g. like in Figure~\ref{fig:turtle_res}. We furthermore see a window that specifies the container to look into every contained resource for the object of a triple that has the \texttt{sosa:resultTime} predicate as a timestamp. All graphs that have timestamps that fall into the last two minutes (see \texttt{ldpsc:logical}) are inside the window (in this case \texttt{:3} and \texttt{:4}) and thus membership triples are added for them. The exact rule executed by the server to create the membership triples can be formalized by the SPARQL CONSTRUCT query in Figure~\ref{fig:rule_logical}. Note that the rule to determine which membership triples are added and consequently which RDF graphs are inside of the window is evaluated every time a GET request reaches the container.

\begin{figure}
\centering
\begin{verbatim}
PREFIX ldpsc: <https://solid.ti.rw.fau.de/public/ns/stream-containers#>
PREFIX ldp: <http://www.w3.org/ns/ldp#>

CONSTRUCT {
    ?memberResource ?memberRelation ?resource .
} WHERE {
    ?container a ldpsc:StreamContainer ;
        ldpsc:window ?window ;
        ldp:contains ?resource .

    ?window ldpsc:logical ?logicalWindow ;
        ldpsc:contentTimestampRelation ?timestampRelation ;
        ldp:hasMemberRelation ?memberRelation ;
        ldp:membershipResource ?memberResource .

    ?resource ?timestampRelation ?timestamp .

    FILTER(?timestamp >= (NOW() - ?logicalWindow))
}
\end{verbatim}
    \caption{Rule for the membership triples for logical windows}
    \label{fig:rule_logical}
\end{figure}

\begin{figure}
    \centering
    \includegraphics[width=\linewidth]{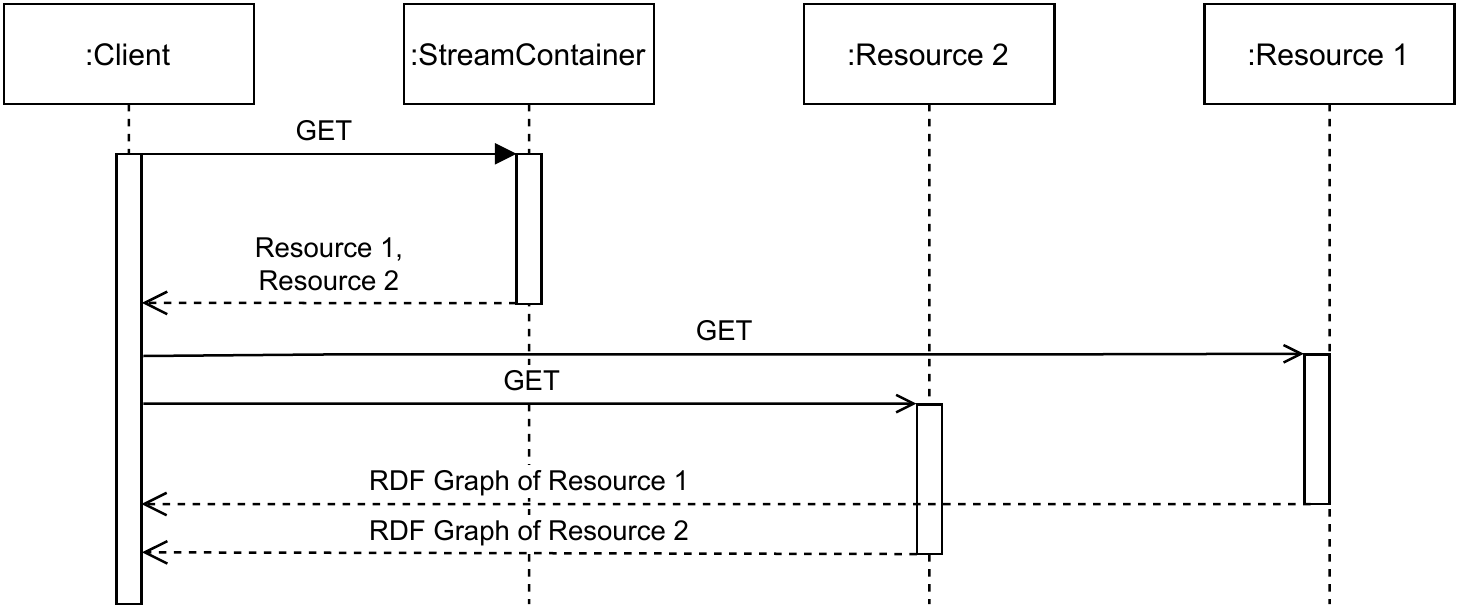}
    \caption{Sequence diagram of a client retrieving the contents of a window}
    \label{fig:sequence}
\end{figure}

A client that wants to retrieve the actual content of the current window thus has to send several GET requests: One to the Stream Container to get the URIs of the resources that are currently in the window and for every returned resource another GET request has to be sent to retrieve the actual RDF graphs. This might seem like a very slow process at first, however all GET requests after the first can be sent asynchronously, so we only need to wait approximately two HTTP round trip times for the results to arrive at the client. An example of the process can be seen in Figure~\ref{fig:sequence}.

The client now has access to all the RDF graphs contained in the window. How exactly those graphs are handled (e.g. whether all graphs are merged into the standard graph, or whether they become named graphs) is up to the client.

To transform the results of the client's processing back into an RDF stream, the client simply has to do a POST request to the respective Stream Container with the RDF graph (including the timestamp) as payload to create a new resource and thus a new data point in the stream. This R2S operator implemented by the Stream Container is commonly called the \texttt{RSTREAM} operator for which in every processing cycle all the results get inserted into the new stream. There however exist two other operators, \texttt{ISTREAM} and \texttt{DSTREAM} that are not directly implemented by the Stream Container. Nevertheless, their behavior can  be reproduced by the client (e.g. by keeping track of what was not in the result in the last cycle and newly added by the current cycle for the \texttt{ISTREAM} operator). Theoretically, it is not a problem to store infinitely many elements in a stream, for practical purposes however, it might be necessary to define a retention policy (e.g. throw away all elements that are older than one day).

\subsection{Semantics}
We now want to define the semantics of our systems comprising Stream Containers and RDF processing clients. We base our notation on the one used in \cite{dell2014rsp} as we want to show the compatibility of our system with the RSP-QL model. Note that for brevity reasons, we focus on the semantics of logical windows as they are the more complex case (compared to physical windows).











Let $t \in T$ be a timestamp where $T$ is the value space of \texttt{xsd:dateTimeStamp}. Usually, the timestamp is taken from the set of natural number for the sake of simplicity but Stream Container use \texttt{xsd:dateTimeStamp} which share an important property with the natural numbers: they are totally ordered.

We define the timestamp extraction function $$t_{pred}(g) = \begin{cases}t\ \ \text{if}\ \ (s, pred, t) \in g\\undefined\ \ \text{else}\end{cases}$$ with $s$ being an RDF term, $pred$ a URI\footnote{In our previous example, $pred$ e.g. was \texttt{sosa:hasTimestamp}.} and $g$ an RDF graph.

An RDF stream $S_{pred}$ is an unbounded sequence of RDF graphs $g_i$ for which $t_{pred}(g_i)$ is defined: $$S_{pred} = (g_1, g_2, g_3,...)$$ Note that we can just use the RDF graph model as it is for this definition -- as opposed to the more common definitions where an RDF stream is a sequence of tuples comprising a graph and a timestamp. We could however easily map our RDF stream definition to the usual one by using the timestamp extraction function: $$S = ((g_1, t_{pred}(g_1)), (g_2, t_{pred}(g_2)),...) = ((g_1, t_1), (g_2, t_2),...) $$

A logical window on a stream $W_{o,c}(S_{pred})$ is defined by the two time instants open $o \in T$ and close $c \in T$. $$W_{o,c}(S_{pred}) = \{g\ |\ g \in S_{pred} \wedge t_{pred}(g) \in (o,c]\}$$

A sliding logical window $\mathds{W}_{\alpha, \beta, t_0}(S)$ is defined by the window size $\alpha \in D$ the step size $\beta \in D$ and the start time instant $t_0 \in T$ with $D$ being the value space of \texttt{xsd:duration}. A logical sliding window uniquely defines a sequence of logical instant windows $$\mathds{W}_{\alpha, \beta, t_0}(S) = (W_{o_0, c_0}, W_{o_1, c_1}, ..., W_{o_i, c_i} ,...)$$
where $o_i = t_0 + i \cdot \beta$ and $c_i = t_0 + \alpha + i \cdot \beta$. $\alpha$ and $\beta$ are usually (in existing RSP systems) given in the query, $t_0$ is implementation specific and $i$ is the index in the stream.

Our implementation of the logical sliding window $\mathds{W}^{SC}_{\alpha}(S_{pred}, t_{eval})$\footnote{$SC$ is not a parameter but short for Stream Container.}, the stream container logical sliding window, just depends on the window size $\alpha$\footnote{The window size $\alpha$ is a duration which is in our implementation specified by \texttt{ldpsc:logical}}. $\beta$ and $t_0$ are determined by the client indirectly by choosing the evaluation time of the sliding window $t_{eval}$. As we implemented a pull architecture and the sliding window is evaluated at every request, $t_{eval}$ is set to the time the request arrives\footnote{Note that $t_{eval}$ is exactly the time determined by the \texttt{NOW()} function in Figure~\ref{fig:rule_logical}}. Thus $$\mathds{W}^{SC}_{\alpha}(S_{pred}, t_{eval}) = W_{t_{eval}, t_{eval} + \alpha}$$

Let $C_{t_0, \beta}$ be a client that is continuously evaluating a sliding window every $\beta$ and starts at $t_0 - \delta$ where $\delta$ is the delay of the HTTP request, then we have a sequence of evaluation times:
$$C_{t_0, \beta} = t_0, t_1, ..., t_i, ...$$ where $t_i = t_0 + i \cdot \beta$.

When a client $C_{t_0, \beta}$ evaluates a logical sliding window on a stream, a sequence of logical windows is determined:
$$C_{t_0, \beta}(\mathds{W}^{SC}_{\alpha}(S_{pred})) = (\mathds{W}^{SC}_{\alpha}(S_{pred}, t_0), \mathds{W}^{SC}_{\alpha}(S_{pred}, t_1), ..., \mathds{W}^{SC}_{\alpha}(S_{pred}, t_i), ...)$$
$$= (W_{o_0, c_0}, W_{o_1, c_1}, ..., W_{o_i, c_i} ,...) = \mathds{W}_{\alpha, \beta, t_0}(S_{pre})$$

The resulting sequence of logical instant windows $(W_1, W_2, W_3,...)$ is the same as for $\mathds{W}_{\alpha, \beta, t_0}(S)$, i.e. we have shown that the behavior of existing RSP systems can also be achieved with our system by timing the clients' requests correctly\footnote{Informally: The \texttt{STEP} parameter is determined by the time between the client's request} and that this behavior is not implementation-specific.






\section{Conclusion \& Future Work}
\label{sec:conclusion}
In this paper, we presented a novel approach for RDF stream processing by decoupling the handling of the stream data including the windowing and streaming operators (S2R and R2S) and the actual processing of static RDF (R2R). We did this by introducing an extension to the Linked Data Platform standard, the Stream Container.

We described the intended overall architecture for Stream Containers to be used in and gave an overview of how the Stream Container is actually implemented. We furthermore provided the formal semantics of the overall system and showed that they are equivalent to already existing RSP systems.

We already have built a working prototype of our proposed architecture\footnote{\href{https://github.com/wintechis/stream-container}{https://github.com/wintechis/stream-container}}, however as future work, we plan to evaluate its performance compared to already existing RSP systems especially in regards to scalability.

All in all, we hope that our proposal will help to encourage the RSP community to better align their systems to the Web architecture, to create a Web of data (streams) where clients can work on the data, process it and create new data, all in a resource-oriented way.

\bibliographystyle{splncs04}
\bibliography{literature}

\end{document}